\documentclass
[preprint,prl,showpacs,showkeys,balancelastpage,onecolumn]{revtex4}%
\usepackage{amsfonts}
\usepackage{amsmath}
\usepackage{amssymb}
\usepackage{graphicx}
\usepackage{revsymb}%
\newtheorem{example}{Example}

\ifx\pdfoutput\relax\let\pdfoutput=\undefined\fi
\newcount\msipdfoutput
\ifx\pdfoutput\undefined\else
\ifcase\pdfoutput\else
\ifx\paperwidth\undefined\else
\ifdim\paperheight=0pt\relax\else\pdfpageheight\paperheight\fi
\ifdim\paperwidth=0pt\relax\else\pdfpagewidth\paperwidth\fi
\fi\fi\fi
\begin{document}
\title[Grover's Algorithm and Diophantine Approximation]{Grover's Quantum Search Algorithm and Diophantine Approximation }
\author{Shahar Dolev}
\author{Itamar Pitowsky}
\affiliation{The Edelstein Center, Levi Building, The Hebrerw University, Givat Ram,
Jerusalem, Israel}
\author{Boaz Tamir}
\affiliation{Department of Philosophy of Science, Bar-Ilan University, Ramat-Gan, Israel.}

\begin{abstract}
In a fundamental paper [\emph{Phys. Rev. Lett. 78}, 325 (1997)] Grover showed
how a quantum computer can find a single marked object in a database of size
$N$ by using only $O(\sqrt{N})$ queries of the oracle that identifies the
object. His result was generalized to the case of finding one object in a
subset of marked elements. We consider the following computational problem: A
subset of marked elements is given whose number of elements is either $M$ or
$K$, our task is to determine which is the case. We show how to solve this
problem with a high probability of success using iterations of Grover's basic
step only, and no other algorithm. Let $m$ be the required number of
iterations; we prove that under certain restrictions on the sizes of $M$ and
$K$ the estimation $m\leq\frac{2\sqrt{N}}{\sqrt{K}-\sqrt{M}}$ obtains. This
bound reproduces previous results based on more elaorate algorithms, and is
known to be optimal up to a constant factor. Our method involves simultaneous
Diophantine approximations, so that Grover's algorithm is conceptualized as an
orbit of an ergodic automorphism of the torus. We comment on situations where
the algorithm may be slow, and note the similarity between these cases and the
problem of small divisors in classical mechanics.

\end{abstract}
\startpage{1}
\maketitle

Consider a database of $N=2^{n}$ elements, which are represented as the basis
vectors of a quantum register $\left\vert a_{j}\right\rangle =\left\vert
a_{1}^{j}\right\rangle \otimes...\otimes\left\vert a_{n}^{j}\right\rangle $,
$a_{k}^{j}\in\{0,1\}$. The state of the register can be any superposition the
basis vectors. Our task is to find in the database one specific element
$\left\vert a_{j}\right\rangle $. At our disposal is an oracle that if given
the required element $\left\vert a_{j}\right\rangle $, will mark it by
rotating its phase by $\pi$. Should the oracle receive a superposition of the
basis elements, it will rotate only the branch of $\left\vert a_{j}%
\right\rangle $. Grover \cite{1} demonstrated that by using $O(\sqrt{N})$
calls to the oracle one can find the marked element $\left\vert a_{j}%
\right\rangle $ with a very high probability. It was also shown \cite{2} that
if one is asked to find any one of $K\ (1<K<N)$ marked elements, it is
possible to reach a high probability of success by calling the oracle
$O(\sqrt{N/K})$ times.

In this paper we consider a variant of the algorithm which can solve fast the
following problem: We know that there is a subset $S$ of marked elements in
the database, and we have an oracle to demarcate them. However, we do not know
exactly how many elements there are in $S$, only that the number is either
$\left\vert S\right\vert =M$ or $\left\vert S\right\vert =K$ for some $0\leq
M<K\leq N/2$. We shall show that under certain restrictions on the values of
$M$ and $K$ and their relations to $N$ we can solve the problem by calling the
oracle $m$ times, where $m<\frac{2\sqrt{N}}{\sqrt{K}-\sqrt{M}}$ (theorem
\nolinebreak1).

Our result reproduces an earlier work of Nayak and Wu \cite{3}, who obtain a
solution to the problem with a probability of success $\geq\frac{2}{3}$ after
calling the oracle $m\leq O(\sqrt{\frac{N}{K-M}}+\frac{\sqrt{M(N-M)}}{K-M})$
times. The authors apply the counting algorithm of Brassard and his
collaborators \cite{4}, which involves the elaborate discrete Fourier
transform \cite{5} on top of Grover's simpler procedure. By contrast, we just
iterate Grover's rotation, and apply simultaneous Diophantine approximations
\cite{6} to calculate the number of iterations that separate the two cases.
Mathematically speaking, this means that we conceptualize Grover's algorithm
as an orbit of a discrete dynamical process on the torus $\mathbb{T}^{2}$.
Note also that all these upper bounds are optimal, up to a constant factor.
This was proved in \cite{3}, see also \cite{7}.

First we shall briefly repeat the algorithm \cite{2} of finding an element of
a set $S\subset\{1,2,...,N\}$, such that $\left\vert S\right\vert =K$. Let us
denote%
\begin{equation}
\left\vert \alpha\right\rangle _{K}\equiv\frac{1}{\sqrt{N-K}}\sum
_{i\not \in S}\left\vert a_{i}\right\rangle \qquad\left\vert \beta
\right\rangle _{K}\equiv\frac{1}{\sqrt{K}}\sum_{i\in S}\left\vert
a_{i}\right\rangle \label{1}%
\end{equation}
Now, write the initial state of the register $\left\vert \psi\right\rangle
=\frac{1}{\sqrt{N}}\sum_{i=0}^{N-1}\left\vert a_{i}\right\rangle $ as a sum of
the two vectors in Eq (\ref{1}) $\left\vert \psi\right\rangle =\sqrt
{\frac{N-K}{N}}\left\vert \alpha\right\rangle _{K}+\sqrt{\frac{K}{N}%
}\left\vert \beta\right\rangle _{K}$, or%
\begin{equation}
\left\vert \psi\right\rangle =\cos\frac{\theta_{K}}{2}\left\vert
\alpha\right\rangle _{K}+\sin\frac{\theta_{K}}{2}\left\vert \beta\right\rangle
_{K},\quad\frac{\theta_{K}}{2}=\sin^{-1}\sqrt{\frac{K}{N}} \label{2}%
\end{equation}
Each step in Grover's algorithm transforms the present state of the register
$\left\vert \psi^{\prime}\right\rangle $ to a new state $G\left\vert
\psi^{\prime}\right\rangle $, where $G$ is a rotation of the plane spanned by
$\left\vert \alpha\right\rangle _{K}$ and $\left\vert \beta\right\rangle _{K}$
by the angle $\theta_{K}$. To perform the rotation we first call the oracle to
reflect $\left\vert \psi^{\prime}\right\rangle $ around $\left\vert
\alpha\right\rangle _{K}$ by introducing a minus sign to the $\left\vert
\beta\right\rangle _{K}$ component; subsequently we reflect the result about
$\left\vert \psi\right\rangle $. After $m$ iterations the state is%
\begin{equation}
G^{m}\left\vert \psi\right\rangle =\cos(m\theta_{K}+\frac{\theta_{K}}%
{2})\left\vert \alpha\right\rangle _{K}+\sin(m\theta_{K}+\frac{\theta_{K}}%
{2})\left\vert \beta\right\rangle _{K} \label{3}%
\end{equation}
All that is left to do is choose $m$ that will bring $G^{m}\left\vert
\psi\right\rangle $ as close as possible to $\left\vert \beta\right\rangle
_{K}$, in other words, we look for an integer $m$ which will satisfy
$\sin(m+\frac{1}{2})\theta_{K}\approx1$. In case $N\gg K$ it follows from the
definition of $\theta_{K}$ that $m$ is of the order of magnitude of
$\sqrt{\frac{N}{K}}$.

In the present problem we are given integers $M<K<N$, and we are told in
advance that there is a subset $S\subset\{0,...,N\}$ of marked elements that
contains either $M$ or $K$ elements but we do not know which is the case. We
wish to find out whether $\left\vert S\right\vert =M$ or $\left\vert
S\right\vert =K$. To find the answer we simply apply Grover's rotations,
without any change. The only difference from Grover's original procedure is
the stopping rule, that is, the number of iterations required before the
measurement is performed. So essentially the same algorithm is solving a
different problem.

We can always represent the initial state $\left\vert \psi\right\rangle $ as
\begin{align}
\left\vert \psi\right\rangle  &  =\cos\frac{\theta_{M}}{2}\left\vert
\alpha\right\rangle _{M}+\sin\frac{\theta_{M}}{2}\left\vert \beta\right\rangle
_{M}\quad if\;\left\vert S\right\vert =M,\label{4}\\
\left\vert \psi\right\rangle  &  =\cos\frac{\theta_{K}}{2}\left\vert
\alpha\right\rangle _{K}+\sin\frac{\theta_{K}}{2}\left\vert \beta\right\rangle
_{K}\quad if\;\left\vert S\right\vert =K\nonumber
\end{align}

With $\sin\frac{\theta_{M}}{2}=\sqrt{\frac{M}{N}}$ and $\sin\frac{\theta_{K}%
}{2}=\sqrt{\frac{K}{N}}$. Our purpose is to compute the number of iterations
$m$ with the following property: If $\left\vert S\right\vert =M$ the rotation
$G$ operates in the plane spanned by $\left\vert \alpha\right\rangle _{M}$ and
$\left\vert \beta\right\rangle _{M}$, and $G^{m}(\left\vert \psi\right\rangle
)$ is close to $\left\vert \alpha\right\rangle _{M}$, the vector of $N-M$
\emph{unmarked} elements; however, if $\left\vert S\right\vert =K$ the
rotation $G$ operates in the plane spanned by $\left\vert \alpha\right\rangle
_{K}$ and $\left\vert \beta\right\rangle _{K}$ while $G^{m}(\left\vert
\psi\right\rangle )$ is close to $\left\vert \beta\right\rangle _{K}$, the
vector of $K$ \emph{marked} elements. We do not know in advance which is the
case, but if such an integer $m$ is found, and $G$ has been iterated $m$
times, all that is left to do is measure the quantum register. If the result
is one of the elements of $S$ (which we check by another query of the oracle)
then it is clear with high probability that $\left\vert S\right\vert =K$ ,
otherwise, $\left\vert S\right\vert =M$. As usual, the probability of success
can be further increased by repeating the process.

The \emph{existence }of such an integer $m$ follows from the theorem of
Kronecker on simultaneous Diophantine approximations \cite{6} : \emph{let
}$\xi_{1},\xi_{2},...,\xi_{r}$\emph{\ be irrational numbers which are linearly
independent over the rationals, and let }$\eta_{1},\eta_{2},..._{,}\eta_{r}%
$\emph{\ be any real numbers, and }$\varepsilon>0$\emph{\ real. Then there are
integers }$p_{1},p_{2},...,p_{r}$\emph{\ and an integer }$l$\emph{\ such that}%

\begin{equation}
\left\vert l\xi_{j}-\eta_{j}-p_{j}\right\vert <\varepsilon\quad j=1,2,...,r
\label{5}%
\end{equation}
In our case $r=2$, and we wish to find an \emph{odd }integer $l=2m+1$ which
approximates $\xi_{1}=\frac{\theta_{M}}{4\pi}$ to $\eta_{1}=0$, and $\xi
_{2}=\frac{\theta_{K}}{4\pi}$ to $\eta_{2}=\frac{1}{4}$ \cite{8}. Only in rare
cases such $\xi_{1}$ or $\xi_{2}$ are rationals, or dependent over the
rationals \cite{9}. The trouble is that it is very hard to obtain a universal
bound on the minimal number $l$ that satisfy Eq (\ref{5}). In the general case
of arbitrary $\xi_{j}$'s and $\eta_{j}$'s no such universal bound exists. In
our more specific case, when we consider all $M$, $K$, and $N$, it is an open problem.

Luckily, there is an interesting range of values of $M$ and $K$ for which a
small odd $l$ does exist.\ Denote $\gamma=\frac{\theta_{K}}{\theta_{M}}%
=\frac{\sin^{-1}(\sqrt{\frac{K}{N}})}{\sin^{-1}(\sqrt{\frac{M}{N}})}$, our
main result is

\textbf{Theorem } \textbf{a}. \emph{If} $M<K<\frac{N}{2}$ \emph{satisfy}
$\sqrt{K}<16(\gamma-1)^{2}\sqrt{N}$ t\emph{hen there are natural numbers} $l$
\emph{and} $p$\emph{, such that} $l$ \emph{is odd, and} $l\leq\frac{4\sqrt{N}%
}{\sqrt{K}-\sqrt{M}}$\emph{, and}
\begin{equation}
\left\vert l(\frac{\theta_{K}}{4\pi})-p-\frac{1}{4}\right\vert <2(\gamma
-1)\qquad\left\vert l(\frac{\theta_{M}}{4\pi})-p\right\vert <(\gamma-1)
\label{6}%
\end{equation}
\textbf{b}. \emph{Let} $\varepsilon>0$ \emph{and consider the cases where}
$K<(1+\frac{\varepsilon}{2\sqrt{2}})^{2}M$. \emph{Then the inequality
}$(\gamma-1)<\frac{\varepsilon}{2}$ \emph{is satisfied, so that }$\left\vert
l(\frac{\theta_{K}}{4\pi})-p-\frac{1}{4}\right\vert <\varepsilon$\emph{, and
also} $\left\vert l(\frac{\theta_{M}}{4\pi})-p\right\vert <\varepsilon$.

The proof of the theorem is given towards the end of the paper. The theorem
allows us to solve the problem with a high probability of success. Suppose
that we have iterated the algorithm $m=\frac{l-1}{2}$ times, $m<\frac
{2\sqrt{N}}{\sqrt{K}-\sqrt{M}}$, and subsequently measured the register. Then
the probability of getting the wrong result is determined in the following
way: First, suppose that there are $K$ elements in $S$, then the probability
of getting an \emph{unmarked} element after the measurement is%
\begin{equation}
\left\vert \left\langle G^{m}\psi|\alpha\right\rangle _{K}\right\vert
^{2}=\cos^{2}(m\theta_{K}+\frac{\theta_{K}}{2})=\cos{}^{2}(\frac{l\theta_{K}%
}{2})<\sin^{2}(2\pi\varepsilon) \label{7}%
\end{equation}
where the last inequality follows from the theorem. Likewise, if there are $M$
elements in $\left\vert S\right\vert $, the probability of measuring a
\emph{marked} element after $m$ iterations is%
\begin{equation}
\left\vert \left\langle G^{m}\psi|\beta\right\rangle _{M}\right\vert ^{2}%
=\sin^{2}(m\theta_{M}+\frac{\theta_{M}}{2})=\sin^{2}(\frac{l\theta_{M}}%
{2})<\sin^{2}(2\pi\varepsilon) \label{8}%
\end{equation}
again, the last inequality follows from the theorem. Note that $\varepsilon$
need not be excessively small. Even if we take $\sin^{2}(2\pi\varepsilon
)=\frac{1}{4}$, that is $\varepsilon=\frac{1}{12}$, then a few repetitions of
the algorithm will give the correct answer with overwhelming probability.

In light of the theorem one can see Grover's algorithm as a discrete dynamical
process on the torus $\mathbb{T}^{2}$: Consider the subset of $\mathbb{T}^{2}$
given by $D=\{(l\frac{\theta_{K}}{4\pi}(\operatorname{mod}1),l\frac{\theta
_{M}}{4\pi}(\operatorname{mod}1))\ \nolinebreak;\nolinebreak\ l\nolinebreak%
\ odd\ \}$. If $\frac{\theta_{K}}{4\pi}$ and $\frac{\theta_{M}}{4\pi}$ are
independent over the rationals, then $D$ is dense in $\mathbb{T}^{2}$; this is
just Kronecker's theorem (with the slight variation that we consider only odd
$l$'s). If $l=1,3,..$ is taken as a discrete time parameter, then $D$ is a
dense orbit of an ergodic dynamical system, and the question is how quickly it
will enter a small prescribed neighborhood of $(\frac{1}{4},0)$. This question
can be generalized to more extensive searches on $\mathbb{T}^{r}$, for
$r\geq3$ (more on this below); or to questions concerning approximations to
other points on the torus, which may be related to the solutions of
Diophantine equations; or finally, to questions regarding continuous rather
than discrete processes, such as adiabatic computations. We shall come back to
this point later.

Here are a few applications of the theorem:

\begin{example}
For $K=M+1$ we need $m=\frac{l-1}{2}<$ $4\sqrt{(M+1)N}$ iterations of Grover's
algorithm to solve the problem up to a probability of error $\sin^{2}%
(2\pi\varepsilon)$. To estimate the range for which this is possible note that
since $\frac{\sin^{-1}(x_{1})}{\sin^{-1}(x_{2})}\geq\frac{x_{1}}{x_{2}}$ for
$0<x_{2}<x_{1}<1$, we have in this case $\gamma-1>\sqrt{1+\frac{1}{M}}%
-1>\frac{1}{3M}$. Therefore, if we choose $\sqrt{\frac{M+1}{N}}<\left(
\frac{4}{3M}\right)  ^{2}$ then the condition of the theorem: $\sqrt
{K}<16(\gamma-1)^{2}\sqrt{N}$ is fulfilled. This means that the range of
application of the algorithm for this case is at least $M\lessapprox
\sqrt[5]{N}$, and the number of steps is $m\leq O(N^{\frac{3}{5}})$.
\end{example}

\begin{example}
For $K=2M$ we can increase the database by adding $rN$ artificial elements,
out of which $rM$ are marked, so they respond positively to the oracle. As a
result we have to separate now between $M^{\prime}=(r+1)M$ and $K^{\prime
}=(r+2)M$, while the total size of the database increases to $N^{\prime
}=(r+1)N$. We proceed as follows: \newline\textbf{a}. The condition that
$K^{\prime}<(1+\frac{\varepsilon}{2\sqrt{2}})^{2}M^{\prime}$ is satisfied if
$r+1>\sqrt{2}\varepsilon^{-1}$. Let $r$ be the minimal integer that satisfies
this inequality. \newline\textbf{b}. Consider the new angles: $\theta
_{M^{\prime}}=\sin^{-1}(\sqrt{\frac{(r+1)M}{(r+1)N}})=\theta_{M}$, and
$\theta_{K^{\prime}}=\sin^{-1}(\sqrt{\frac{(r+2)M}{(r+1)N}})$. Then
$\gamma^{\prime}-1=\frac{\theta_{K^{\prime}}}{\theta_{M^{\prime}}}-1\geq
\sqrt{\frac{(r+2)}{(r+1)}}-1>\frac{\varepsilon}{3\sqrt{2}}$. (We are using
once more the fact that $\frac{\sin^{-1}(x_{1})}{\sin^{-1}(x_{2})}\geq
\frac{x_{1}}{x_{2}}$ for $0<x_{2}<x_{1}<1$, and the minimality of $r$ from
\textbf{a}). \newline\textbf{c}. Consequently, if we assume $\sqrt{\frac{M}%
{N}}<(\frac{2\varepsilon}{3})^{2}$ then the condition $\sqrt{\frac
{K^{^{\prime}}}{N^{^{\prime}}}}=\sqrt{\frac{(r+2)M}{(r+1)N}}<\frac{8}%
{9}\varepsilon^{2}<16(\gamma^{\prime}-1)^{2}$ is fulfilled.\newline\textbf{d}.
Now, apply the theorem to separate between $M$ and $2M$ with an error
$\varepsilon$ in a number of steps $m$ less than $\frac{2\sqrt{N^{\prime}}%
}{\sqrt{K^{\prime}}-\sqrt{M^{\prime}}}<5(r+1)\sqrt{\frac{N}{M}}\approx
5\sqrt{2}\varepsilon^{-1}\sqrt{\frac{N}{M}}$. Since $\varepsilon>\frac{3}%
{2}\left(  \frac{M}{N}\right)  ^{\frac{1}{4}}$ we conclude that we need $m\leq
O\left[  \left(  \frac{N}{M}\right)  ^{\frac{3}{4}}\right]  $ steps for the
separation between $M$ and $2M$, provided $\sqrt{M}<(\frac{2\varepsilon}%
{3})^{2}\sqrt{N}$. The same technique will also work for the case $K=aM$, with
some $a>1$.
\end{example}

\begin{example}
Note that if the conditions of the theorem hold for the triple $M$, $K$, $N$
they also hold for $nM$, $nK$, $nN$ where $n$ is any integer. Also, the angles
$\theta_{M}$, and $\theta_{K}$ remain the same, and therefore so does the
number of iterations required to complete the job, despite the fact that the
database has increased $n$-fold. This means that we should take care only of
triples $M$, $K$, $N$ that do not have a common divisor.
\end{example}

\begin{example}
Suppose that our information is that one of the following cases obtains:
$\left\vert S\right\vert =M_{1}$, or $\left\vert S\right\vert =M_{2}$, or
...,$\left\vert S\right\vert =M_{r}$. We can inductively use multiple
Diophantine approximations as in Eq (\ref{5}): First find an odd integer $l$
such that $\sin(l\frac{\theta_{M_{j}}}{2})\approx1$ for $1\leq j\leq
\lbrack\frac{r}{2}]$, while $\sin(l\frac{\theta_{M_{j}}}{2})\approx0$, for
$[\frac{r}{2}]<j\leq1$. If a measurement discovers a marked element then with
high probability $\left\vert S\right\vert =M_{j}$ for some $1\leq j\leq
\lbrack\frac{r}{2}]$, otherwise it is one of the other cases. Now, divide the
resulting set of possibilities into two halves and continue the process. After
$\sim\log_{2}r$ successive Diophantine approximations we are guaranteed to
find the answer. The trouble is that the larger $r$ is the larger $l$ is
likely to be, and it is not clear when the process yields better than
classical outcomes.
\end{example}

\begin{example}
If $M=0$ and $K>0$, then we are just back with Grover's type algorithm. If
$\left\vert S\right\vert =0$ nothing happens to $\left\vert \psi\right\rangle
=\left\vert \alpha\right\rangle _{M}$, and if $\left\vert S\right\vert =K$ we
get close to $\left\vert \beta\right\rangle _{K}$.
\end{example}

Two remarks on the general problem should be made: Firstly, for arbitrary
values of $M$ and $K$ the minimal size of the number of iterations $m$ depends
on our ability to obtain a \emph{lower bound} on the uniform Diophantine
approximation to the quotient $\gamma=\frac{\theta_{K}}{\theta_{M}}$. By this
we mean finding natural numbers $p$ and $q$ such that $\left\vert
p\gamma-q\right\vert $ is small, \emph{but not too small}. The reason will
become clear from the proof below. Intuitively, if $p\frac{\theta_{K}}%
{\theta_{M}}$ stays close to an integer for a long segment of values of $p$,
then the two numbers $\frac{\theta_{M}}{4\pi}$ and $\frac{\theta_{K}}{4\pi}$
become hard to separate with a small $l$. This means that the general
separation problem runs into a difficulty similar to the problem of small
denominators (or divisors) in classical mechanics \cite{10}. It is likely that
a formulation of the algorithm in terms of a continuous adiabatic quantum
computer will demonstrate more clearly the relation between our problem and
the KAM-type of problems, in the sense that small divisors may show up in the
spectral gap. Note also that some of these small divisor problems may be
overcome by using the trick in \textbf{Example 2}, namely by adding artificial
elements to the database and changing the values of $\theta_{K}$ and
$\theta_{M}$.

Secondly, a remark about the actual value of $l=l(M,K,N)$, the number of
iterations needed to complete the task. The proof below is giving a pretty
good estimation of $l$. However, note that this is essentially a different
problem. Once a "table" of the values of $l$ is generated for the appropriate
$M,K,N$, it can serve all search problems, no matter what the nature of the
objects in the database, and the character of the oracle. Such "table" may
allow us to decide what is the best strategy to use. We shall just briefly
indicate how to formulate this problem algebraically: Denote $T_{l}%
(x)=\cos[l\cos^{-1}(x)]$, then $T_{l}$ is the $l$ degree Chebyshev's
polynomial (of the first kind) \cite{11}. Using Eq (\ref{3}) we see
that\ $\cos(l\frac{\theta_{M}}{2})=T_{l}(\sqrt{\frac{N-M}{N}})$ and similarly
$\cos(l\frac{\theta_{K}}{2})=T_{l}(\sqrt{\frac{N-K}{N}})$. To get rid of the
square roots we can use the identity $2T_{l}^{2}(x)-1=T_{2l}(x)=T_{l}%
(2x^{2}-1)$; so that finally our task is to find the smallest odd $l$ such
that $T_{l}(\frac{N-2M}{N})\approx+1$ while $T_{l}(\frac{N-2K}{N})\approx-1$.
This observation may also assist in generalizing our result to other values of
$M$, $K$, and $N$.

\textbf{Proof of the theorem }: Denote $\gamma=\frac{\theta_{K}}{\theta_{M}%
}=\frac{\sin^{-1}(\sqrt{\frac{K}{N}})}{\sin^{-1}(\sqrt{\frac{M}{N}})}$, we
take the following three steps

\textbf{Step 1:} If $M<K<\frac{N}{2}$ then%

\begin{equation}
0<\gamma-1<\sqrt{2}(\sqrt{K/M}-1) \label{9}%
\end{equation}

That $\gamma>1$ is obvious since $M<K$ and $\sin^{-1}(x)$ is increasing. For
the right hand estimation we use the mean value theorem. First note that if
$0<x_{2}<x_{1}<1$ then%

\[
\frac{\sin^{-1}(x_{1})}{\sin^{-1}(x_{2})}=1+\frac{\sin^{-1}(x_{1})-\sin
^{-1}(x_{2})}{\sin^{-1}(x_{2})-\sin^{-1}(0)}=1+\frac{\sqrt{1-x_{3}^{2}}}%
{\sqrt{1-x_{4}^{2}}}\frac{x_{1}-x_{2}}{x_{2}}%
\]
for some $x_{3}$ and $x_{4}$ such that $x_{1}>x_{4}>x_{2}>x_{3}>0$. Now,
substitute $x_{1}=\sqrt{K/N}$ and $x_{2}=\sqrt{M/N}$, and remember that
$1-x_{3}^{2}<1,$ and $x_{4}^{2}<\frac{K}{N}$, so that the condition
$K<\frac{N}{2}$ entails $\gamma<1+\sqrt{2}(\sqrt{K/M}-1)$.

\textbf{Step 2:} Choose $p$ to be the nearest \emph{odd} integer to $\frac
{1}{4(\gamma-1)}$, then $\left\vert p-\frac{1}{4(\gamma-1)}\right\vert \leq1$,
and also $p\leq\frac{1}{2}(\gamma-1)^{-1}$ (assuming $\gamma-1\leq\frac{1}{4}%
$). and altogether:
\begin{equation}
\left\vert p\gamma-p-\frac{1}{4}\right\vert \leq(\gamma-1)\qquad p\leq\frac
{1}{2(\gamma-1)}\qquad p\ odd \label{10}%
\end{equation}
Now, add the condition $\sqrt{\frac{K}{N}}<16(\gamma-1)^{2}$. Denote by $s$
the nearest \emph{odd} integer to $\frac{4\pi}{\theta_{M}}$, then $\left\vert
s-\frac{4\pi}{\theta_{M}}\right\vert \leq1$, and put $l=ps$ therefore $l$ is
also odd. Then, since $\sin^{-1}(x)\leq\frac{\pi}{2}x$ for $0\leq x\leq1$, we
have from Eq \nolinebreak(\ref{10})
\begin{align}
\left\vert l\frac{\theta_{K}}{4\pi}-p-\frac{1}{4}\right\vert  &
\leq\left\vert p\frac{\theta_{K}}{\theta_{M}}-p-\frac{1}{4}\right\vert
+\frac{\theta_{K}}{4\pi}p\left\vert s-\frac{4\pi}{\theta_{M}}\right\vert
<2(\gamma-1)\label{11}\\
\left\vert l\frac{\theta_{M}}{4\pi}-p\right\vert  &  =p\frac{\theta_{M}}{4\pi
}\left\vert s-\frac{4\pi}{\theta_{M}}\right\vert <(\gamma-1)\nonumber
\end{align}

\textbf{Step 3:} Let $\varepsilon>0$. To complete the proof all we have to do
is impose the condition $2(\gamma-1)\leq\varepsilon$. But then by Eq (\ref{9})
this will be satisfied if $\sqrt{K}\leq(1+\frac{\varepsilon}{2\sqrt{2}}%
)\sqrt{M}$. To estimate $l$ note that by our definition $p\approx\frac
{1}{4(\gamma-1)}$ while $s\approx\frac{4\pi}{\theta_{M}}$ where $\approx$
indicates equality up to $\pm1$. Hence, $l=ps\approx\frac{\pi}{\theta
_{K}-\theta_{M}}$. Using once more the mean value theorem for $\sin^{-1}(x)$
we get $l\leq\frac{4\sqrt{N}}{\sqrt{K}-\sqrt{M}}$. $\blacksquare$

Returning to the issue of small denominators, consider how it is avoided in
our proof: On the one hand $\gamma-1$ is small, indeed $\gamma-1<\frac
{\varepsilon}{2}$ is our basic constraint. On the other hand $p$ is of the
order of magnitude of $(\gamma-1)^{-1}$, and $l=ps>p$, so that $\gamma-1$
cannot be too small. This is the balance that should be struck if we wish to
generalize the result to other values of $\gamma=\frac{\theta_{K}}{\theta_{M}%
}$; we have to obtain a uniform Diophantine approximation $\left\vert
p\gamma-q\right\vert $ which is small, but reasonably bounded from below. A
further complication is that $p$ has to be odd. General lower bounds of this
kind exist for algebraic numbers, but $\gamma$ is typically transcendental.
However, we are dealing with a very special case for which a good lower bound
may exist. Also, we can move from a bad case to a better one by adding
artificial elements to the database, as in \textbf{Example 2}.

\textbf{Conclusion} Given an oracle that identifies the elements of a subset
$S\subset\{1,2,...,N\}$, and knowledge that either $\left\vert S\right\vert
=M$ or $\left\vert S\right\vert =K$, for $M<K$, we demonstrated how to decide
which is the case by iterating Grover's rotation $m\leq\frac{2\sqrt{N}}%
{\sqrt{K}-\sqrt{M}}$ times. The algorithm is working for a certain range of
values $M$, $K$, and $N$, and employs simultaneous Diophantine approximations.
This means that we conceive of Grover's algorithm as an orbit of an ergodic
automorphism of the torus $\mathbb{T}^{2}$, and ask how quickly it enters a
given open subset of $\mathbb{T}^{2}$. We showed how to apply this process in
some special cases, and noted that in other cases the algorithm may be
frustrated because of a `small divisor'\ type of problem.

\textbf{Acknowledgements} We thank Michael Ben-Or and Scott Aaronson for
calling our attention to earlier work on the subject. One of us (IP) is
grateful for the support of the Israel Science Foundation grant number 879/02.


\begin{thebibliography}{99}
\bibitem {8}Kronecker's theorem remains valid when restricted to odd integers
$l$. To see that apply Kronecker's theorem to find integers $n$ and $q_{j}$
such that $\left\vert n\xi_{j}-\frac{1}{2}(\eta_{j}-\xi_{j})-q_{j}\right\vert
<\frac{\varepsilon}{2}$, and then multiply the inequalities by $2$.
\end{thebibliography}
\end{document}